\documentclass[a4paper]{jpconf}

\usepackage{graphicx}

\begin{document}
\title{ Initial Design of a High-Power Ka-Band Klystron}

\author{M. Behtouei\textsuperscript{1}, L. Faillace\textsuperscript{2}, M. Ferrario \textsuperscript{1}, B. Spataro \textsuperscript{1}, and A. Variola \textsuperscript{1}}

\address{$^1$INFN, Laboratori Nazionali di Frascati, P.O. Box 13,
I-00044 Frascati, Italy\\ $^2$INFN - Sezione di Milano, Via Celoria, 16 - 20133 Milano, Italy  }

\ead{Mostafa.Behtouei@lnf.infn.it}

\begin{abstract}
Accelerating structures operating in Ka-Band are foreseen to achieve gradients around 150 MV/m. Among possible applications of a Ka-Band accelerating structure we refer to the beam phase-space manipulation for the Compact Light XLS project as well as medical and industrial applications. In this paper, a Ka-Band Klystron amplifier is being investigated in order to feed Ka-Band accelerating structures. The initial design is presented including the high-power DC gun and the beam focusing channel.
\end{abstract}

\section{Introduction}
 In the framework of the Compact light XLS project, the main linac frequency is F=11.994GHz. In order to compensate the non-linearity distortions due to the RF curvature of the accelerating cavities, the use of a compact third harmonic accelerating structure working at F = 35.982 GHz is required \cite{behtouei2019ka,faillace2019compact,behtouei2019new,PhdDissertation}. Our concern is to design a high power Ka band klystron in order to feed  a constant impedance accelerating structure operating on the $2\pi/3$ mode with  an average (100-125) MV/m  accelerating electric field range by using the conservative main RF parameters. For this reason  we are planning to finalize the structure design as well as engineering of the RF power source that will be able to produce up to a 40-50 MW input power by using a SLED system \cite{ivanov2009active,ivanov2013active} since the theoretical efficiency of the third harmonic Klystron operating on the $TM_{01}$ mode is around the $18\%$ (about a factor 3 less than the standard klystron efficiency).  As a result, in order to obtain the RF power source requested, a 100 MW electron gun beam power is needed by knowing that the output cavity is operated in the third harmonic of the drive frequency in the $TM_{01}$ mode. Space charge force is one of the limitations which doesn't allow to have identical velocity for each accelerated electrons after passing through the cavities by affecting the bunching process which leads to the low efficiency. The key element to control and measure such a force is known as perveance, $K=I\ V^{-3/2}$. The higher the perveance, the stronger the space charge and consequently the weaker the bunching. On the other hand, since the perveance means how much current comes out of cathode for a certain voltage difference applied between the cathode and anode, to have a higher beam current we should rise the perveance, but higher perveance leads to low efficiency. As a result, we have to find an optimal perveance to maintain a good efficiency. In this paper, a Ka-Band Klystron amplifier is being investigated in order to feed Ka-Band accelerating structures. The initial design is presented including the high-power DC gun and the beam focusing channel.

\section{Electron Gun Injector Beam Dynamics Estimations}
We have started the design of a Pierce-type electron gun as injector of the klystron operating at Ka-Band (35 GHz) in order to feed the accelerating structure. In this case, the cathode-anode voltage is about 500 kV, producing a beam current of about 200 A and beam power up to 100 MW. In Fig. 1, the preliminary simulation of the electron gun with CST is shown. 
\begin{figure}[h]
\begin{center}
\begin{minipage}{38pc}
\includegraphics[width=10pc]{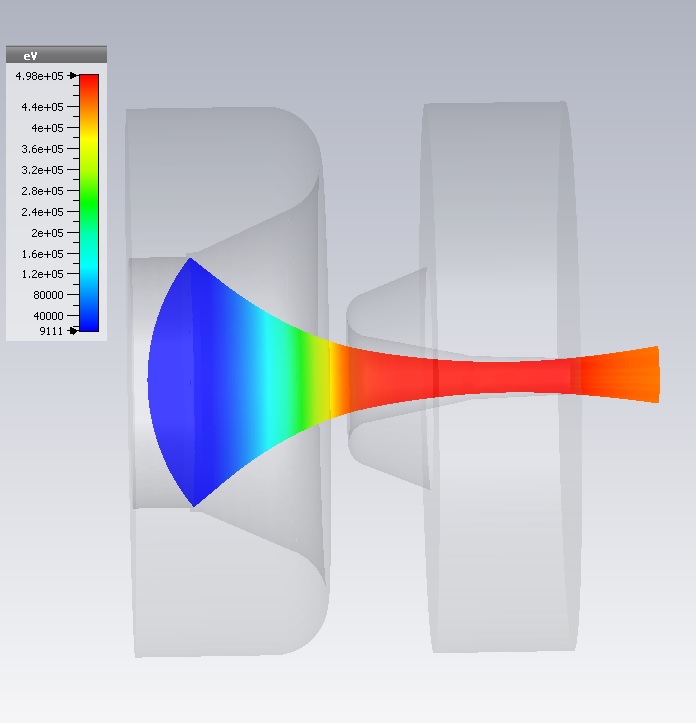}\hspace{6pc}%
\includegraphics[width=10pc]{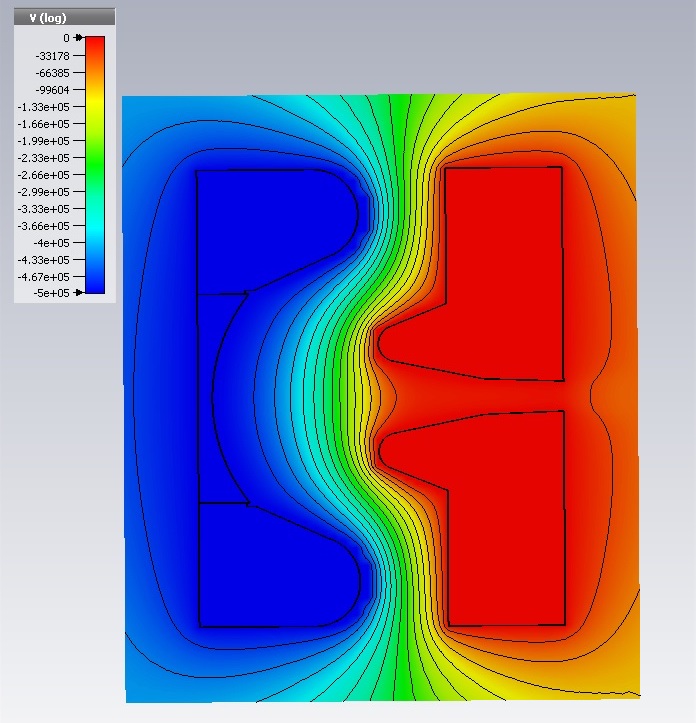}
\caption{\footnotesize{Preliminary electron gun design from CST. Beam trajectory (left) and Equipotential lines (right) are shown.}}
\end{minipage} 
\end{center}
\end{figure}

Beam trajectory (left) and electric field equipotential lines (right) are shown. The cathode-anode geometry was optimized to adjust the electric field equipotential lines in order to obtain maximum beam current extraction and capture (above 200 A). Design parameters of the diode gun for the Ka-band klystron are shown in Table 1.

\begin{table}[h]

\caption{Design parameters of diode gun for Ka-band klystron}

\begin{center}
\small{ \begin{tabular}{lll}
\br
Design parameters&  \\ 
 \mr
Beam power [MW]&  118\\  
Beam voltage [kV]& 500\\ 
Beam current [A]  & 238\\ 
 $\mu-$ perveance $[I/V^{3/2}]$&0.67 \\ 
 Cathode diameter [mm]& 76 \\ 
 Max EF on focusing electrode [kV/cm]& 240 \\ 
 Electrostatic compression ratio& 210\\ 
 \br
\end{tabular}}
\end{center}
\end{table}

We obtained the electrostatic beam compression ratio of 210: 1. To rise the compression ratio we have to apply a focusing magnetic field and it will be reported in the next section. The $\mu$ perveance of the device is 0.67 $A V^{-3/2}$. It is common to use micro perveance because its order is typically of $10^{-6}$ $A V^{-3/2}$. Maximum electric field on focusing electrode is about 240 $kV/cm$ which is a reasonable value. In order to avoid possible damage for a safety operation margin in terms of pulse length, the RF windows, power supply hardware stability, etc., we have decided to work with a 480 kV cathode-anode voltage with maximum electric field of $\sim$ 200 kV/cm on the focusing electrode as it will be reported in the next section.

\section{Magnetostatic Simulation}

As the beam propagates through the beam pipe after the electron gun exit, the transverse dimension begins to increase due to the intense space charge, especially for the high current beam of several hundreds of amperes. This is the reason why we have to use a transverse focusing magnet. To achieve the required compression of the beam after existing the electron gun, a solenoid with two different distributions has been used. Two different field profiles are presented in the Fig.  2 (right) and corresponding beam envelopes are shown (left).

\begin{figure}[h]{Magnetostatic Simulation by CST Particle Studio}

\begin{minipage}{18.5pc}
\includegraphics[width=17.5pc]{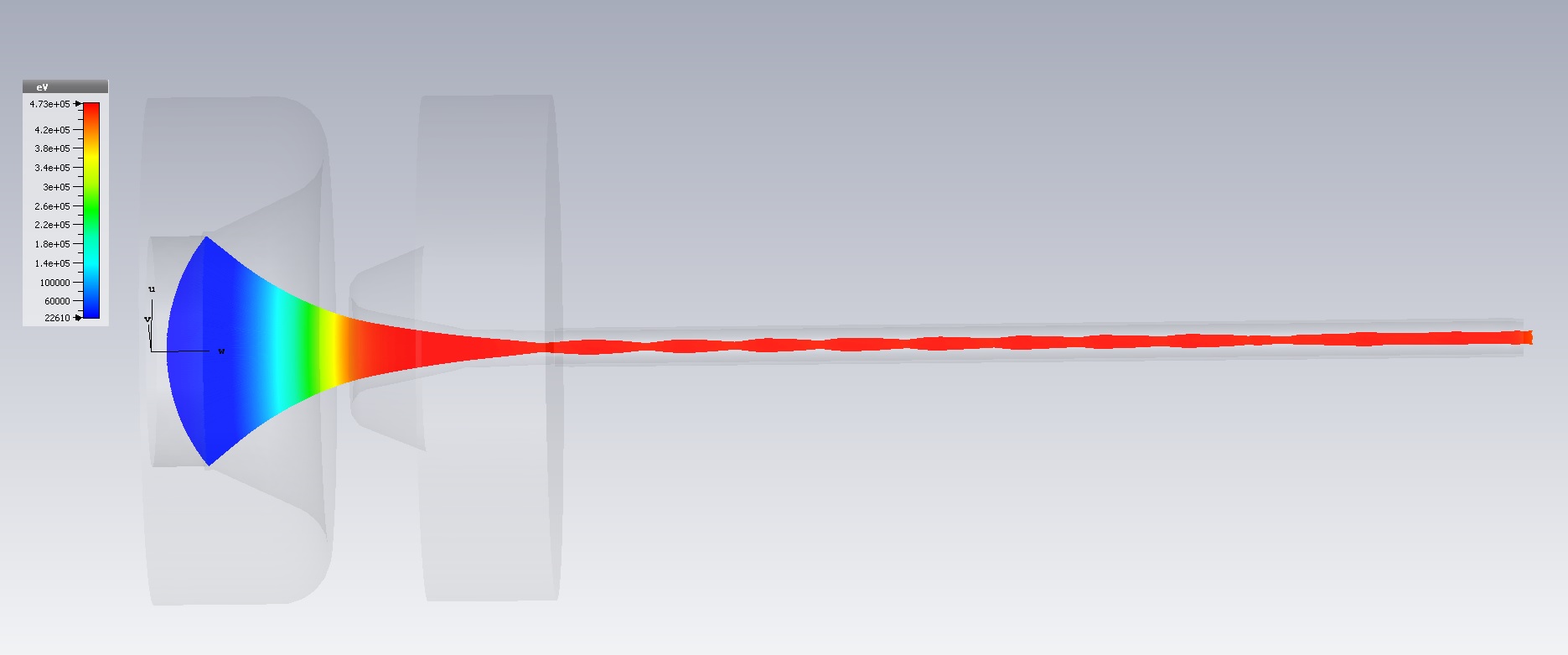} Model 1\\
\includegraphics[width=17.5pc]{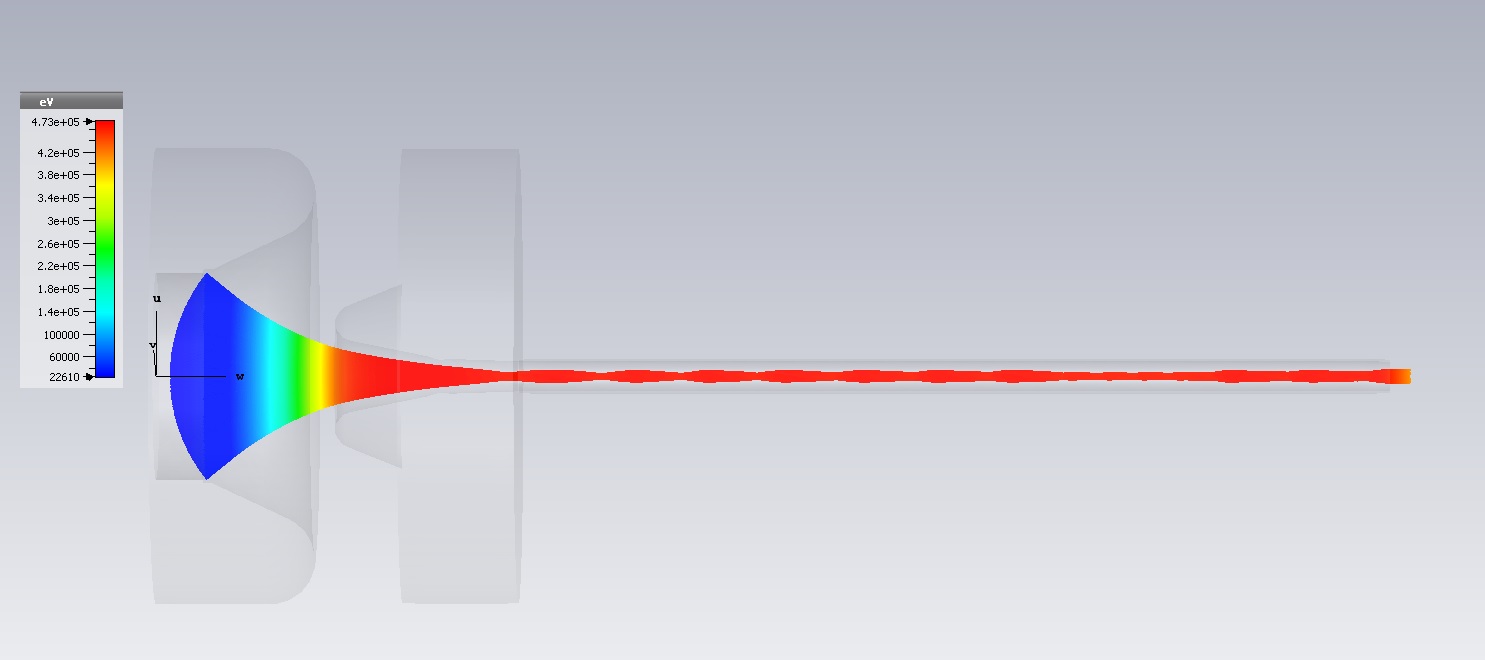} \ Model 2

\end{minipage}
\begin{minipage}{18.5pc}
\includegraphics[width=20pc]{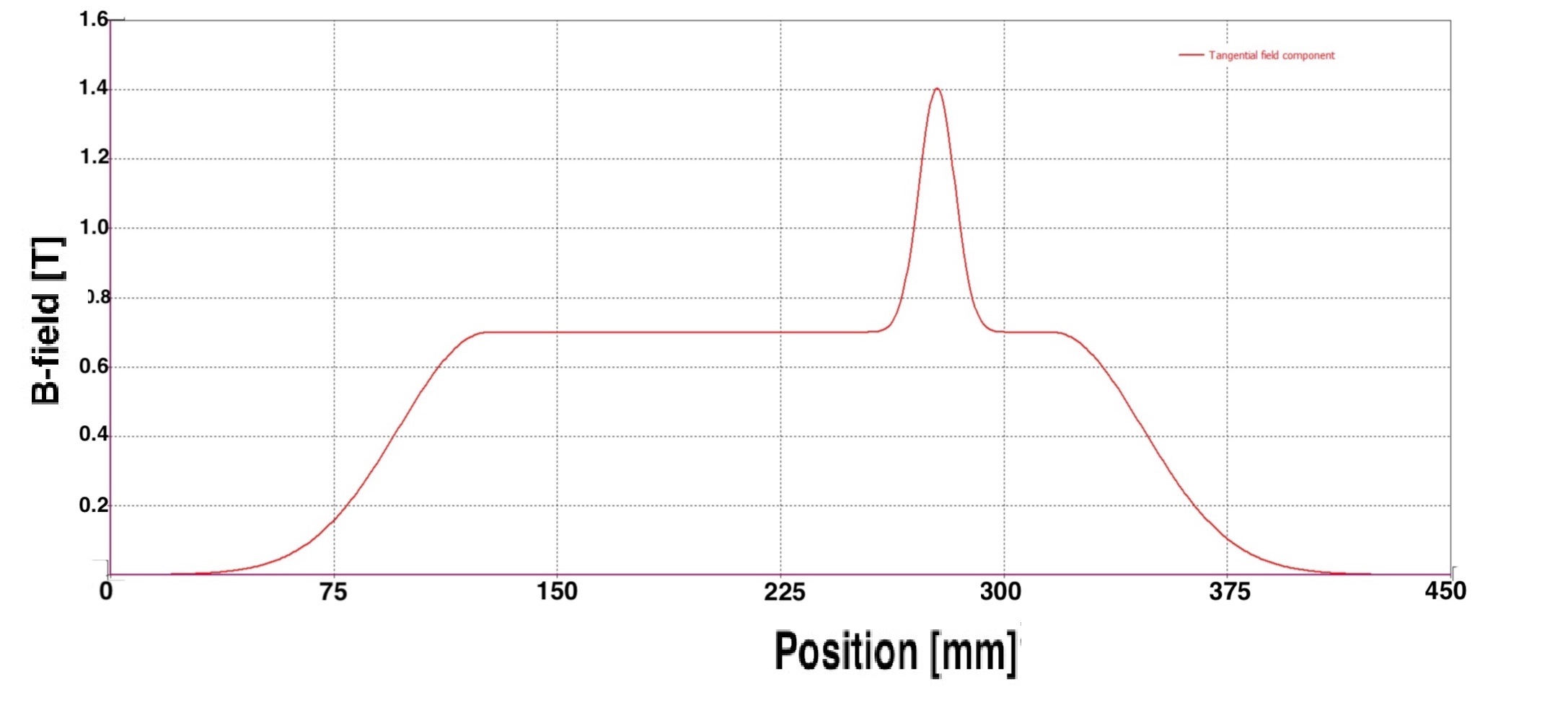}\\
\includegraphics[width=20pc]{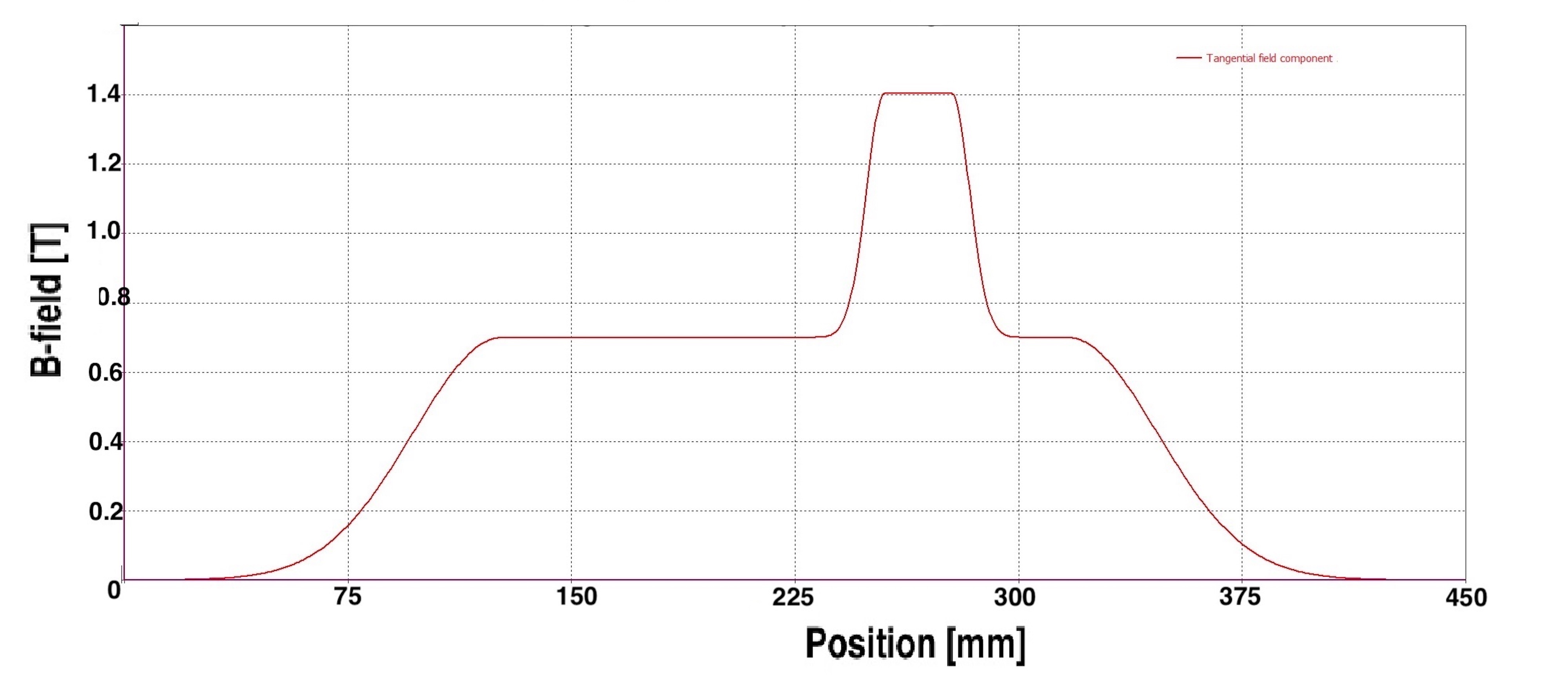}
\end{minipage} 
\caption{\footnotesize{Beam trajectory along the propagation direction (left) and axial magnetic field distribution (right) for two different models. Model 1 has a small peak of 14 kG model 2 has the same amplitude but a constant magnetic field for a distance of about 30 mm at that magnitude.}}
\end{figure}

Model 1 has a small peak of 14 kG and model 2 has the same amplitude but a constant magnetic field of 14 kG along a distance of 30 mm in order to have a narrow beam radius for the purpose of inserting the output cavities. These cavities are operated in Ka-Band and therefore they require a small beam radius of about 2 mm. The gun parameters are presented in Table 2.

\begin{table}[h]

\caption{Design parameters of the gun with focusing magnetic field along the beam axis}

 \begin{center}
\begin{tabular}{llll} 
 \br
Design parameters& Model 1& Model 2\\ 
 \mr
Beam power [MW]&  104&104\\ 
Beam voltage [kV]& 480&480\\  
Beam current [A]  & 218&218\\ 
 $\mu-$ perveance $[I/V^{3/2}]$&0.657 &0.657\\  
 Cathode diameter [mm]& 76& 76 \\ 
  Pulse duration [$\mu$ sec]& 1& 1 \\ 
Minimum beam radius in magnetic system [mm]& 1.04&1.09\\ 
 Max EF on focusing electrode [kV/cm]& 208& 208 \\ 
 Electrostatic compression ratio& 210:1&210:1\\ 
Beam compression ratio& 1635:1&1489:1\\ 
 Emission cathode current density [$A/cm^2$]& 3.92&3.92\\ 
Transverse Emittance of the beam [mrad-cm]& 1.39 $\pi$&1.41 $\pi$\\ 
Beam energy density [kJ/$cm^2$]&   5.37&5.37\\ 
 \br
\end{tabular}
\end{center}
\end{table}

 In the region where magnetic field is 7 kG, the beam radius is $\sim$ 2.2 mm  and considerably higher than Brillouin limit which is 0.6 mm. Likewise for the region where the field is 14 kG, the beam radius is $\sim$ 1 mm which again is much bigger than the Brillouin limit which is about 0.3 mm.

As it can be observed from Fig. 2, for the first rise and the last decay of the magnetic field profile we use the Rician distribution whose equation is as follows \cite{rice1945mathematical}:

\begin{equation}
\small{f (z | \nu, \sigma)=\frac{z}{\sigma^2}\  e^ {(\frac{-(z^2+\nu^2)}{2\sigma^2})}\ I_0 (\frac{z\nu}{\sigma^2})}
\end{equation}

where $I_0 (z)$ is the modified Bessel function of the first kind with order zero. The parameter $\nu$ is the position of the center of the peak, $\sigma$ (the standard deviation) controls the width of the Rician distribution function and  z is the height of the curve's peak. The reason we use this kind of distribution is that it is similar to the practical distribution used in the solenoids.

As the perveance means how much current comes out of cathode for a certain voltage difference applied between the cathode and anode, to have a high beam current we should rise the perveance, but higher perveance leads to low efficiency \cite{chin2008design} and we have to find an optimal perveance to maintain a good efficiency. In our case the $\mu$-perveance is 0.657 $A/V^{3/2}$ which is a common perveance for designing a modern klystron.

To avoid of voltage breakdown and limitations of cathode loading, the maximum possible beam compression is necessary for designing the device \cite{yakovlev1999limitations}. In order to increase the beam compression one should take into account the transverse emittance, indeed, by increasing the beam compression, allowing the minimum beam radius, the transverse emittance rises as we can observe from Fig 3.

\begin{figure}[h]
\begin{center}
\includegraphics[width=31.5pc]{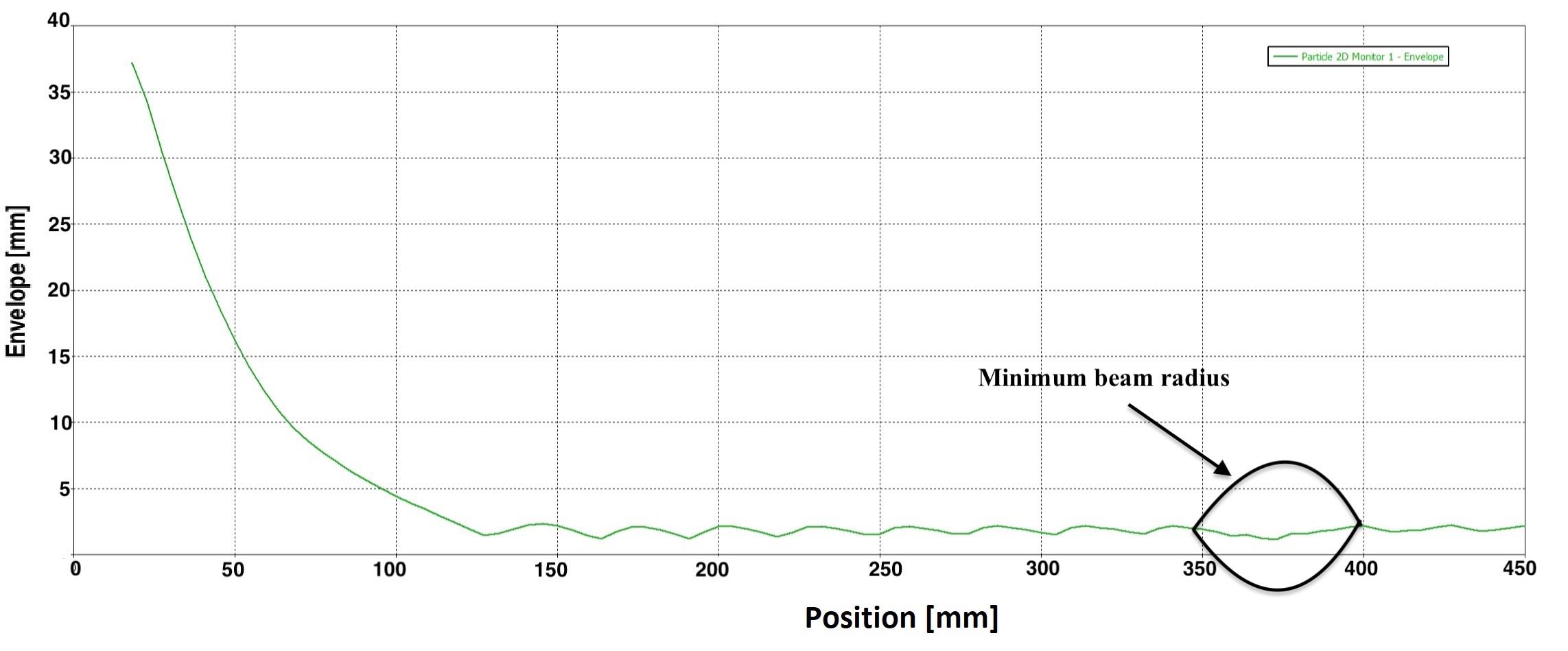}(a)
 \includegraphics[width=31.5pc]{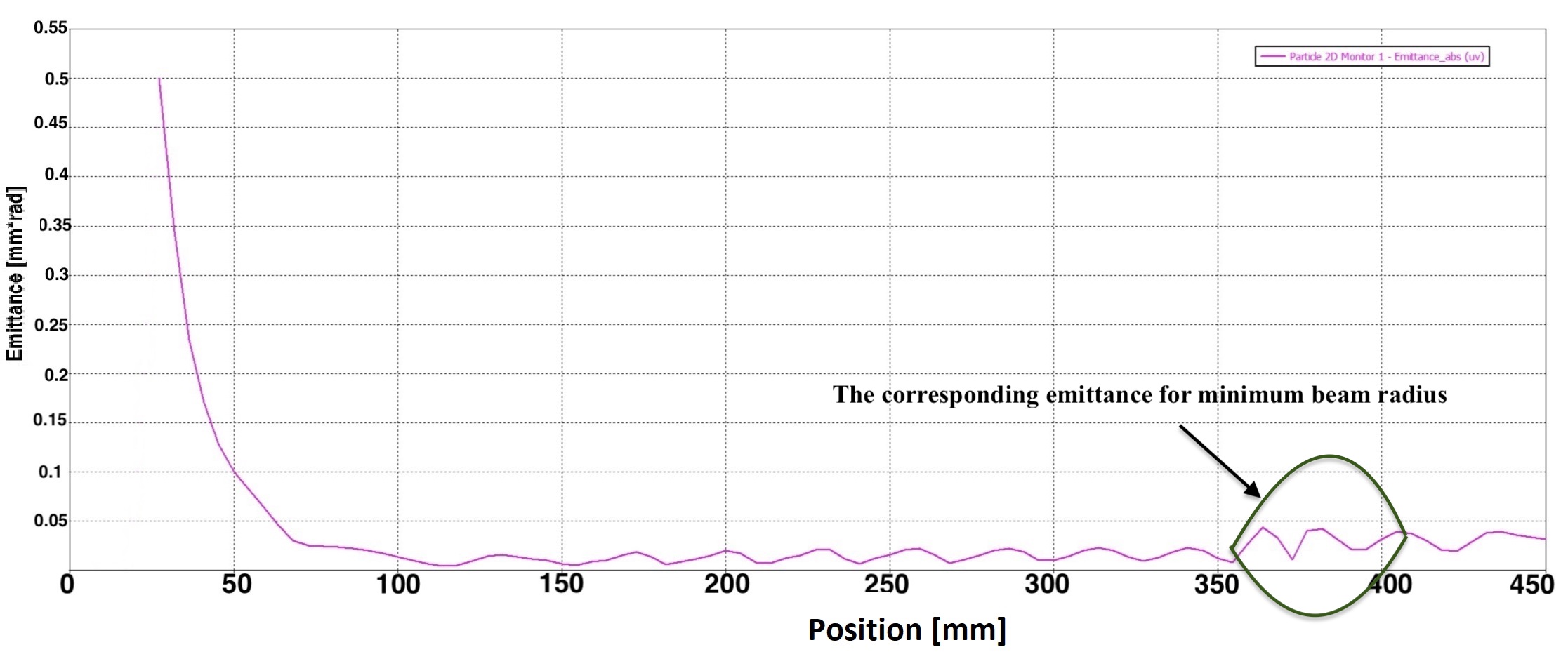}(b)
\caption{\footnotesize a) beam envelope for the Model 2 in which there is a peak magnetic field of 14 kG for a distance of about 30 mm. The minimum beam radius can be observed from the figure and b) transverse emittance of the beam along the beam axis at the presence of the focusing magnetic field. The transverse emittance rises in the region we have the minimum beam radius.}
\end{center}
\end{figure}

We have obtained the magnetostatic beam compression ratio of 1635:1 for the model 1 where the beam radius is $\sim$ 1 mm. It should be noted that it would be possible to rise the beam compression ratio to more than 2000:1 just by decreasing the beam radius to 0.9 mm. The maximum possible compression ratio is 4914 where the beam radius arrives to the Brillouin limit of 0.6 mm. The transverse emittances of the beam for the model 1 and 2 are 1.39 $\pi$ (m\ rad-cm) and 1.41 $\pi$ (m\ rad-cm), respectively.

\section{The main device limitations}

In designing a high power klystron we have some limitations: a) beam current   b) beam radius  and c) cathode material. As we have mentioned before, in order to have a high beam current we should rise the perveance, but higher perveance leads to low efficiency and we have to find an optimal perveance to maintain a good efficiency. The beam radius r cannot be less than the Brillouin limit, 

\begin{equation}
\small{r_b=\frac{0.369}{B}\ \sqrt{\frac{I}{\beta\gamma} }\ mm}
\end{equation}

where, I is the beam current (I=235A), $\beta$ denotes v/c for relativistic particle ($\beta=0.860$), $\gamma$ stands for the relativistic mass (energy) factor  ($\gamma=1.957$) and B is magnetic field in kG ( B=14 kG). Finally we investigated the limitation which is related to the cathode material used as the source of current emission. Tungsten filament and Lanthanum hexaboride (LaB$_6$) are two common materials used as source of current emission. LaB$_6$ has bigger lifetime than Tungsten. The other advantage of LaB$_6$ is that the emitted current is much bigger due to the low work function. We reported the properties of these material in Table 3.

\begin{table}[h]

\caption{ Properties of cathode materials}

\begin{center}
\begin{tabular}{llll} 
 \br
Properties&Tungsten filament & Lanthanum hexaboride (LaB6)\\ 
 \mr
Operating temperature [$^\circ K$]&  2700-3000 &1700-2100 \\ 
Emitted current ($J_c$) [$A/cm^2$]& 1.75 & 40-100 \\  
Required vacuum [Pa]& $10^{-3}$ &$10^{-4}$\\ 
Average life time [hr]& 60-100 & longer than Tungsten\\ 
 Work function [eV] & 4.5 & 2.7\\ 
 \br
\end{tabular}
\end{center}

\end{table}

We decided to work with LaB$_6$ as the cathode material in space charge limited regime in order to get a greater current emission and less cathode damage. 

\section{Analytical method for estimating the dimensions of electron gun device}      
      
An expression for the potential distribution between the cathode and anode may be obtained from considering Poisson's equation. Poisson's equation in spherical coordinates is,

\begin{equation}
\small{\frac{1}{r^2} \frac{\partial}{\partial r} (r^2 \frac{\partial V}{\partial r})+\frac{1}{r^2 sin \theta}\frac{\partial}{\partial \theta} (sin \theta \frac{\partial V}{\partial \theta})+\frac{1}{r^2 sin^2\theta}\frac{\partial^2V}{\partial \phi^2}=-\frac{\rho}{\epsilon_0}}
\end{equation}

  We have no variation of the potential in $\theta$ and $\phi$ coordinates because of the symmetry about the axes and the equation above becomes:

\begin{equation}
\small{\frac{1}{r^2} \frac{\partial}{\partial r} (r^2 \frac{\partial V}{\partial r})=-\frac{\rho}{\epsilon_0}=\frac{I}{4\pi r^2 \nu \epsilon_0}}
\end{equation}
 
where $\nu$ is the electron velocity. The above equation can be solved in terms of a series by H. M. Mott-Smith method. The final solution takes the form \cite{langmuir1913effect,langmuir1924currents},

\begin{equation}
\small{I=\frac{16\pi \epsilon_0}{9} \sqrt{\frac{-2e}{m}} \frac{V^{3/2}}{(-\alpha)^2}}
\end{equation}

where m stands for the electron mass, $\alpha$ is a function of the ratio of the radii $r_a$ and $r_c$ of the spheres in which these radii are equivalent to  the anode radius and the radius of the emitter, respectively  (see Fig. 4) \cite{pierce1940rectilinear}, $\gamma=log(\frac{R_a}{R_c})=log(\frac{r_a}{r_c}$) and 

\begin{figure}[h]
\includegraphics[width=20pc]{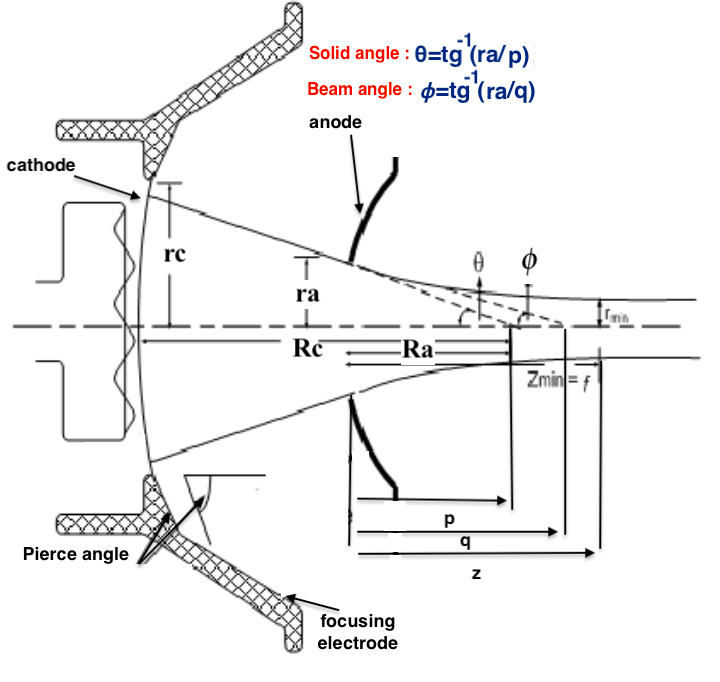}(b)\hspace{1pc}%
\includegraphics[width=18pc]{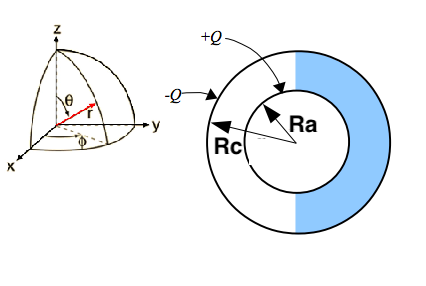} (a)
\caption{a) Schematic view of the Pierce-type gun geometry  b) Two concentric conducting spheres of inner and outer radii $R_a$ and $R_c$, equivalent with the DC gun}
\end{figure}

 \begin{equation}
\small{\alpha= \gamma-0.3 \gamma^2+0.075\gamma^3-0.0143\gamma^4+....}
\end{equation}

 For klystrons in the space-charge limit, the beam current is proportional to the klystron
voltage raised to the three-halves power. The constant of proportionality is known as the perveance.

 \begin{table}[h]
\caption{Comparison between analytical and numerical results for estimating the dimensions of electron gun device}
\begin{center}
\begin{tabular}{llll} 
 \br
Parameters& Analytical& Numerical (CST Particle Studio)\\ 
 \mr
$\frac{r_c}{r_a}$&  2.67&3.78\\ 
$r_a$ [mm]&13.85&9.79 \\ 
Solid angle, $\theta$& 34.72$^\circ$&26.08$^\circ$\\ 
Beam angle, $\phi=tg^{-1} (r_a/q)$&28.04$^\circ$ &18.07$^\circ$\\ 
 Beam current [A]  & 228&238\\ [0.5ex] 
 \br
\end{tabular}
\end{center}
\end{table}

The beam angle $\phi$ (see Fig. 4a) can be obtained from the electrostatic lens effect due to the anode aperture ($\phi=tg^{-1}(r_a/q)$) \cite{haimson1962some}. From analogy between light optics and charged-particle optics we have $\frac{1}{p}+\frac{1}{q}=\frac{1}{f}=\frac{E_2-E_1}{4V_a}$ where $p=R_a$ and $E_2=0$ is the field on the anode. Then we have [13],

\begin{equation}
\small{\frac{1}{q}=\frac{1}{R_a}-\frac{E_1}{4V_a}}
\end{equation}

where E$_1$ is the field on the cathode side of the anode and V$_a$ stands for the voltage. The comparison between analytical and numerical results for estimating the dimensions of electron gun device is presented in Table 4. As we observe from the table, we obtained emitted currents  of 238 A and 228 A from the cathode by numerical and analytical methods, respectively. The analytical estimation of the solid angle, $\theta$, is larger comparing to numerical results and this results in a bigger anode radius as we observe from the Table 4.  Numerical calculations for the other parameters such as beam angle and the ratio between cathode radius and anode radius are in agreement with the analytical ones. 

\section{Conclusions}

In this paper, we have performed the initial electromagnetic and beam dynamics design of an RF Klystron amplifier in order to feed Ka-Band accelerating structures, by using the Microwave CST code. The klystron works on the third harmonic of the bunched electron beam ($\sim$35 GHz). The electron flow is generated from a high-voltage DC gun (up to 500 kV) and the cathode-anode geometry was optimized to adjust the electric field equipotential lines in order to obtain maximum beam current extraction and capture (above 200 A). The electron beam is then transported through the klystron channel. The beam confinement is obtained by means of a high magnetic field produced by superconducting coils, in the current design , which was analytically imported into the code. The channel optimization allows to deliver a 100 MW electron beam with a spot size below 2mm diameter. 

We are currently working on the 2D beam dynamics design of the input, bunching and output RF cavities of this klystron and further details will be given in a following paper. A tapered tunnel is expected to be installed in order to allocate Ka-band output cavities. We are also considering the possibility of using normal conducting coils instead of superconducting ones. 

\section*{Acknowledgement}

This work was performed in the framework of the Compact Light XLS project. Useful and constructive discussions with Prof. Jay Hirshfield and Dr. Sergey Shchelkunov are gratefully acknowledged.

\section*{References}

\bibliography{KaBandKlystron}
\bibliographystyle{iopart-num}

\end{document}